# A STATISTICAL ANALYSIS OF THE PERFORMANCE OF FIRST YEAR ENGINEERING STUDENTS AT UNSW CANBERRA AND THE IMPACT OF THE STATE WHERE THEY UNDERTOOK YEAR 12 STUDY

John F. Arnold and Leesa A. Sidhu
UNSW Canberra
UNSW Australia

**Abstract**

UNSW Canberra at the Australian Defence Force Academy is a unique institution in Australia as it attracts its undergraduate students from all Australian states and territories more or less in accord with the distribution of the Australian population. Each course at UNSW Canberra is then made up of a cohort of students who have undertaken secondary education in the different states and territories but who, at university, are undertaking the same course with the same assessment as one another. This allows some comparison to be made as to how the various state and territory secondary education systems have prepared them for their tertiary study at UNSW Canberra.

In this paper we conduct a preliminary analysis of the performance of UNSW Canberra engineering students in the first year, first semester courses Engineering Mathematics 1A and Engineering Physics 1A. The results obtained thus far demonstrate that while there is little difference in performance between students from most states and territories, performance of students from one state is well below that of the others.

## 1. Introduction

Despite its relatively small population, Australia runs seven separate education systems for its primary and secondary students within its six states and two territories. Most Australian universities draw local students primarily from the state in which they are located and so can be expected to tailor their programs to meet the knowledge base of these students.

UNSW Canberra at the Australian Defence Force Academy draws its students from all states and territories across Australia, more or less in proportion to the distribution of the Australian population. A particular course at UNSW Canberra is therefore made up of a cohort of students who have come from most, if not all, of the states and territories. This gives a unique opportunity to analyse how well each of the secondary education systems has prepared students for tertiary study at UNSW Canberra.

This is not an area that has been widely reported in the past. In 1998, Catchpole and Anderson [1] reported a comparison between the performance of New South Wales and Victorian students in first year Mathematics at UNSW Canberra and noted that students from New South Wales significantly outperformed their Victorian colleagues with the same tertiary entrance rank over the period 1995 to 1997. In their analysis of first year mathematics results at UNSW Canberra in 2006, Barry and Chapman [2] found that the student scores in a diagnostic test (which tested pre-calculus mathematics) were a better predictor of success in first year mathematics than tertiary entrance rank alone. They also found that students having the same tertiary entrance rank scored worse in first year mathematics if they had studied intermediate rather than advanced mathematics in senior or if they had studied in Queensland rather than in another state or territory.

In this paper, we consider students from all seven secondary education systems within Australia. Our analysis is based upon student results obtained in first semester first year for engineering students entering UNSW Canberra between 2007 and 2014 in the subjects Engineering Mathematics 1A and Engineering Physics 1A. Prior to analysis, the data set was carefully scrutinised to ensure that it only included students who had an ATAR and an Australian state or territory recorded as their place of residence, and who had completed Year 12 in the year immediately before beginning study at UNSW Canberra or in the previous two years, provided that they had not studied at another tertiary institution in the interim. The latter group includes all Navy students studying at UNSW Canberra who undertake a full time year of Navy training prior to coming to UNSW Canberra as well as students in the other services who choose to take a gap year prior to attending UNSW Canberra.

Table 1 shows the numbers of students undertaking Engineering Mathematics 1A and Engineering Physics 1A by state and territory. The students from South Australia and Northern Territory have been combined as the Northern Territory Certificate of Education and Training is based on the South Australian Certificate of Education noting that the Northern Territory cohort is extremely small.

Table 1 Number of Students Undertaking Engineering Mathematics 1A and Engineering Physics 1A in 2007-2014

| State or Territory | Number of Students | |
|---|---|---|
| | Engineering Mathematics 1A | Engineering Physics 1A |
| Australian Capital Territory | 60 | 59 |
| New South Wales | 204 | 203 |
| Queensland | 286 | 286 |
| South Australia & Northern Territory | 82 | 80 |
| Tasmania | 23 | 24 |
| Victoria | 124 | 125 |
| Western Australia | 33 | 33 |
| Total | 812 | 810 |

## 2. Results

### 2.1 Engineering Mathematics 1A

Figure 1 shows the results obtained by students in Engineering Mathematics 1A versus their Year 12 ATAR. Students from different states are indicated by a different colour. Figure 2 shows the same results as Figure 1 but for only the three states with the largest cohorts of students (New South Wales, Queensland and Victoria).

Examining Figure 1 and Figure 2, it is clear that there are a large number of Queensland students in the bottom right quarter of the data set compared to students from other states and territories. This would seem to indicate that students from Queensland are achieving lower academic results in first year engineering mathematics than students from other states and territories with the same ATAR.

**Figure 1 Engineering Mathematics 1A Results versus ATAR with Individual State Marking – all States and Territories**

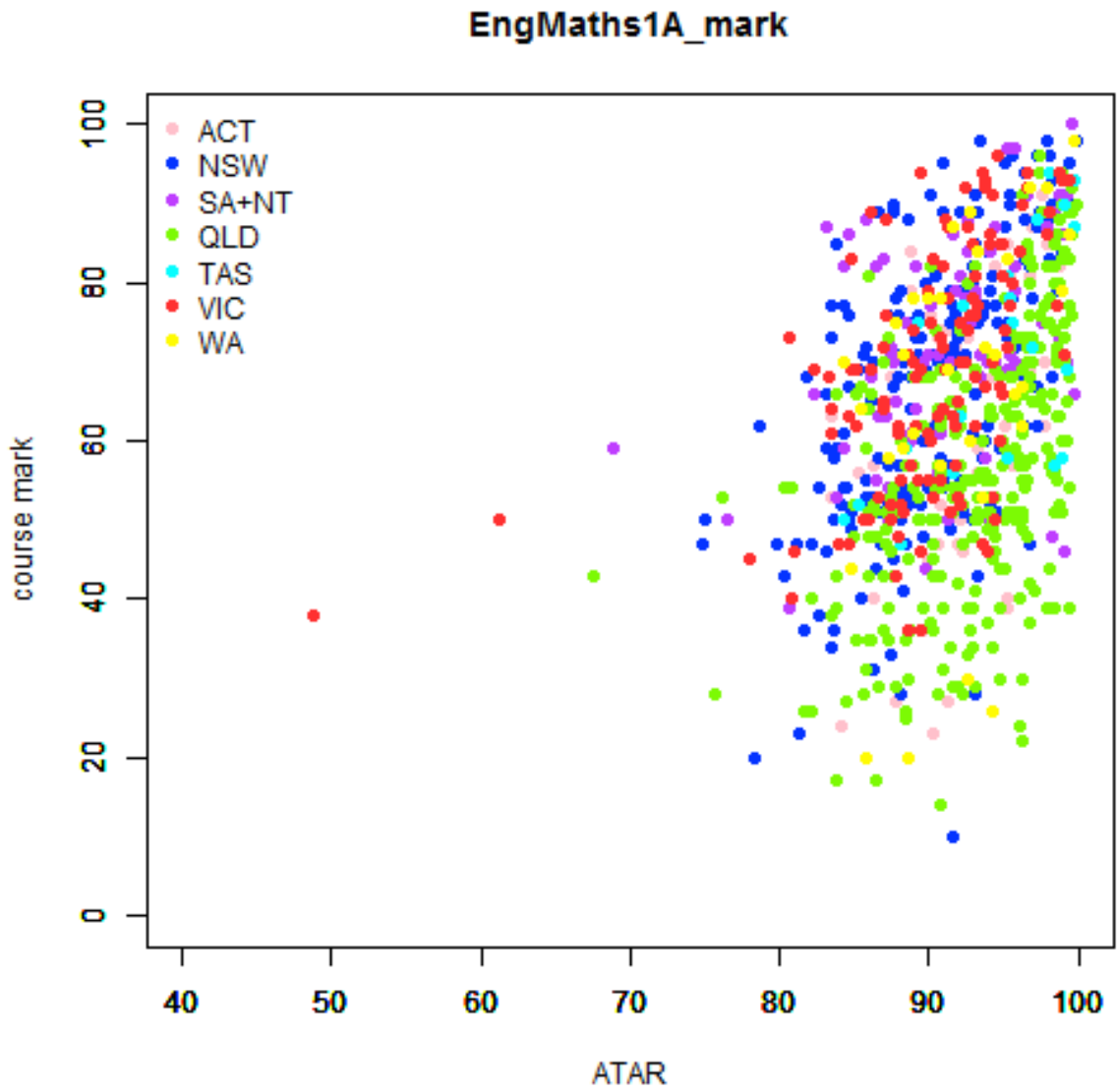

**Figure 2 Engineering Mathematics 1A Results versus ATAR with Individual State Marking – New South Wales, Queensland and Victoria**

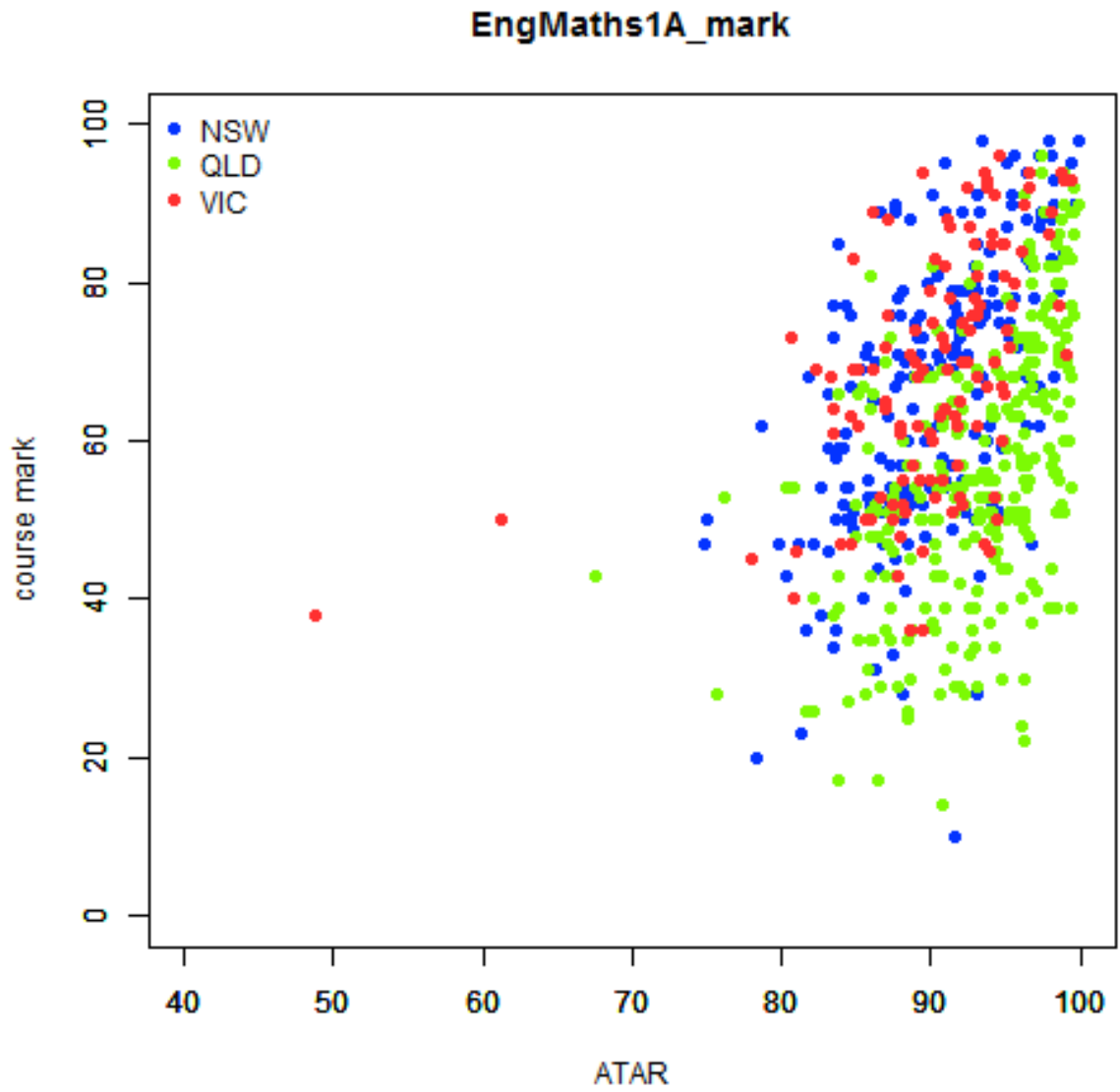

Table 2 summarises the Engineering Mathematics 1A results obtained by students from each state and territory.

Table 2 Mean and Standard Deviation of Results in Engineering Mathematics 1A by State and Territory over the Period 2007-2014.

| State or Territory | Average Mark | Standard Deviation |
|---|---|---|
| Australian Capital Territory | 64.0 | 17.8 |
| New South Wales | 66.5 | 17.3 |
| Queensland | 57.0 | 16.9 |
| South Australia & Northern Territory | 71.1 | 14.3 |
| Tasmania | 69.3 | 14.8 |
| Victoria | 68.8 | 15.3 |
| Western Australia | 66.6 | 20.3 |

In order to explore this further, we undertook a one-way ANOVA to determine whether the mean Engineering Mathematics 1A results across the seven educational areas were significantly different. The ANOVA analysis indicates that the hypothesis of equality of mean Engineering Mathematics 1A results across all states and territories should be rejected ($p = 5 \times 10^{-15}$). Pairwise comparisons of the mean Engineering Mathematics 1A results for the states and territories indicate that the pairs of mean results are significantly different (at the 5% significance level) for Queensland and every other state and territory with the exception of the Australian Capital Territory ($p=0.052$). However, when Queensland is excluded, none of the other differences between the states and territories are statistically significant.

## 2.2. Engineering Physics 1A

Figure 3 shows the results obtained by students in Engineering Physics 1A versus their Year 12 ATAR. Students from different states are indicated by a different colour. Figure 4 shows the same results as Figure 3 but for only the three states with the largest cohorts of students (New South Wales, Queensland and Victoria).

Examining Figure 3 and Figure 4 there is a similar pattern as with Engineering Mathematics 1A in that there are a large number of Queensland students in the bottom right quarter of the data set compared to students from other states and territories. This would again seem to indicate that students from Queensland are achieving lower academic results in Engineering Physics 1A than students from other states and territories with the same ATAR.

**Figure 3 Engineering Physics 1A Results versus ATAR with Individual State Marking – all States and Territories**

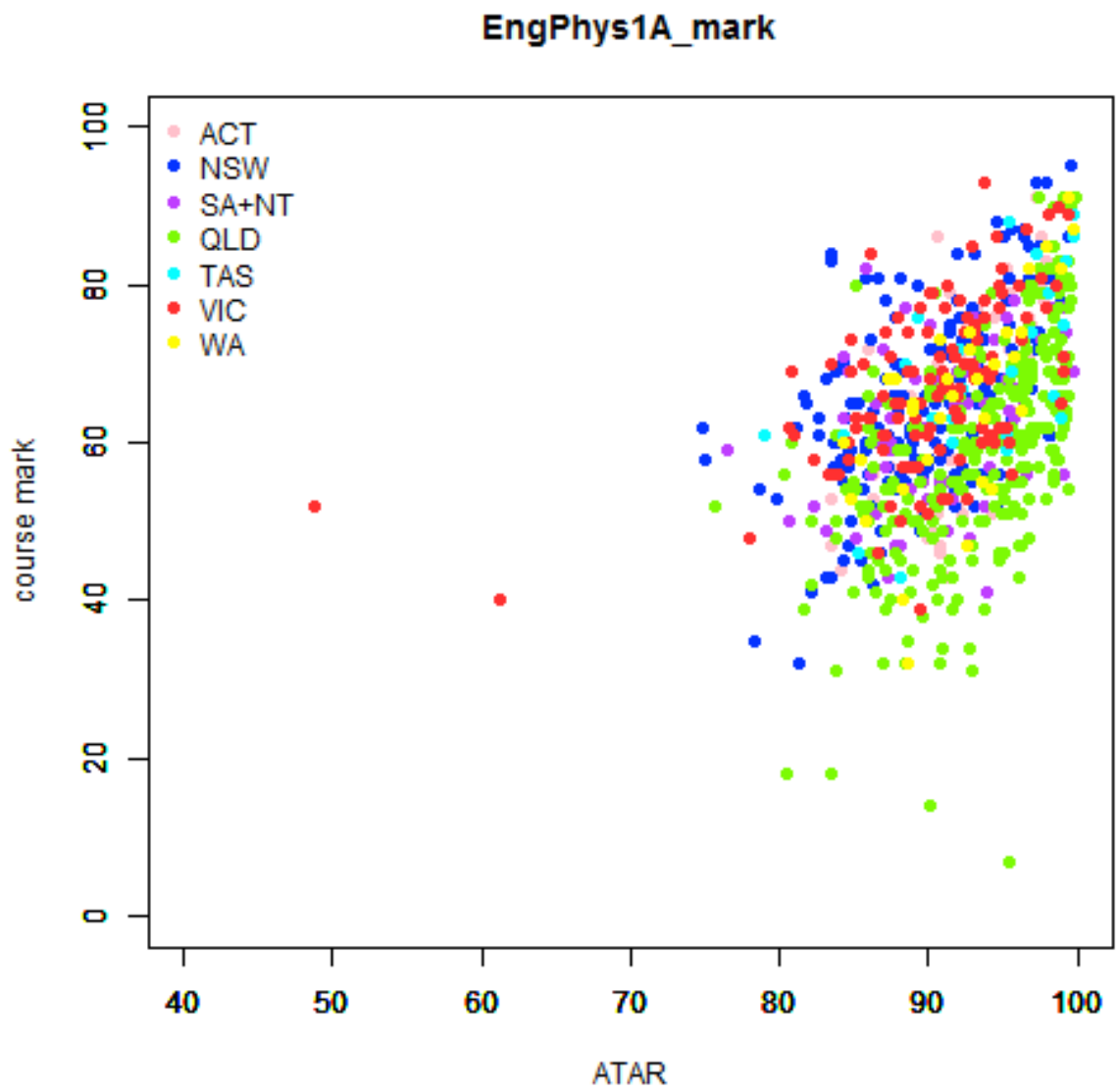

**Figure 4 Engineering Physics 1A Results versus ATAR with Individual State Marking – New South Wales, Queensland and Victoria**

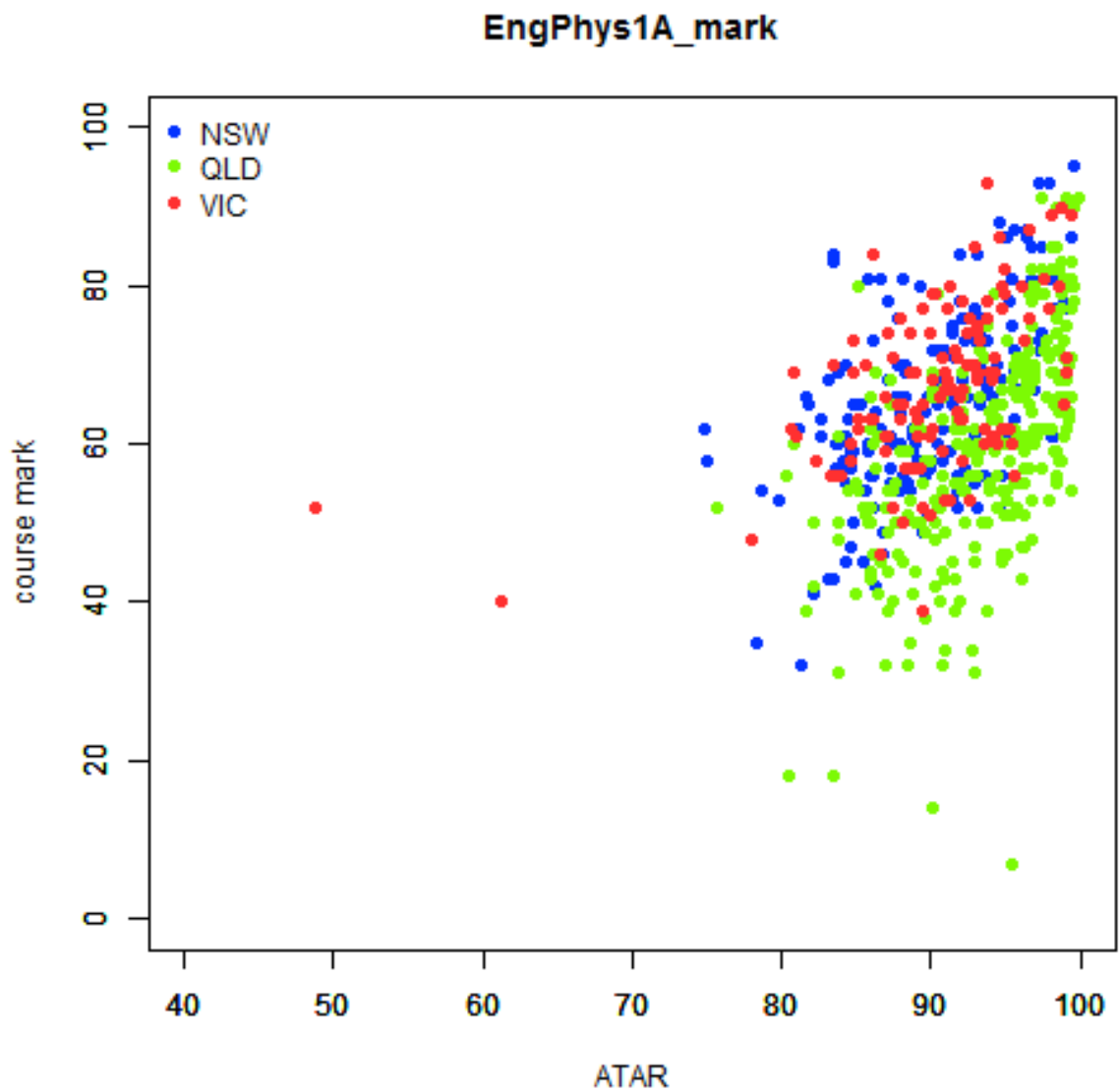

Table 3 summarises the Engineering Physics 1A results obtained by students from each state and territory.

Table 3 Mean and Standard Deviation of Results in Engineering Physics 1A by State and Territory over the Period 2007-2014.

| State or Territory | Average Mark | Standard Deviation |
|---|---|---|
| Australian Capital Territory | 64.5 | 12.5 |
| New South Wales | 65.0 | 11.4 |
| Queensland | 59.7 | 13.4 |
| South Australia & Northern Territory | 63.4 | 10.3 |
| Tasmania | 69.1 | 12.1 |
| Victoria | 67.1 | 10.2 |
| Western Australia | 65.2 | 13.1 |

In order to explore this further, we undertook a one-way ANOVA to determine whether the mean Engineering Physics 1A results across the seven educational areas were significantly different. The ANOVA analysis indicates that the hypothesis of equality of mean Engineering Mathematics 1A results across all states and territories should be rejected ($p=1 \times 10^{-8}$). Pairwise comparisons of the mean Engineering Physics 1A results for the states and territories indicate that the pairs of mean results are significantly different (at the 1% significance level) for Queensland and three other states and territories (New South Wales, Victoria and Tasmania). Again when Queensland is excluded, none of the other differences between the states and territories are statistically significant.

## 4. Conclusion

UNSW Canberra is unique in the Australian tertiary education system in that it takes in undergraduate students from all around Australia in rough proportion to the distribution of the Australian population. Results achieved during the first year of undergraduate studies allow some conclusions to be drawn as to how well each of the primary and secondary education systems in Australia have prepared students for study at UNSW Canberra.

The study reported in this paper considers the performance of students in the courses Engineering Mathematics 1A and Engineering Physics 1A undertaken during their first semester of study at UNSW Canberra. The preliminary results presented indicate that the performances of students from all states and territories in these subjects are not the same. Further analysis reveals that the reason for this is that Queensland students perform significantly less well than students from almost all other states and territories in Engineering Mathematics 1A. In Engineering Physics 1A students from Queensland perform statistically significantly worse than students from three other states. For both Engineering Mathematics 1A and Engineering Physics 1A there are no statistically significant differences between the performances of students from states other than Queensland.

There is no reason to believe that the academic ability of Queensland students should be any different from the academic ability of students from other states and territories. One possible explanation of the difference is that UNSW Canberra is attracting students of lower ability from Queensland compared to other areas of Australia. However, we have considered the mean ATAR scores from each jurisdiction. Results indicate that while the mean ATAR scores across the states and territories are significantly different, the differences between the mean ATAR scores of Queensland students joining UNSW Canberra are either not statistically significant or significantly higher than students from these other jurisdictions.

The observed difference between the performances of Queensland students and those from other states and territories could be due to Queensland students being younger than many of their counterparts. However, Queensland students at UNSW Canberra have not always performed poorly in Mathematics relative to their colleagues. An analysis of results for students studying first year Mathematics at UNSW Canberra from 1986 to 1996 showed that, although there was a statistically significant difference between the mean results across the states and territories ($p<1 \times 10^{-3}$), Queensland students did not perform significantly worse than students from most other states and territories. This suggests that there may be other reasons for the relatively poor performance of Queensland students in the current analysis.

Another possibility is that there is a significant difference between the education system in Queensland and the education systems in other states and territories. We are not qualified to answer this question but would be keen to collaborate with other experts to better understand this issue. We acknowledge that Queensland students may be learning important skills that are not being assessed in Engineering Mathematics 1A and Engineering Physics 1A. However, the content covered and the assessment methods in our courses are similar to those of many other Go8 universities and our Engineering degree program has been accredited by Engineers Australia.

A more detailed analysis will follow. Future work will model the student results in Engineering Mathematics 1A and Engineering Physics 1A as depending on the factors state or territory, the year in which the student undertook the course, socio economic status and ATAR.